\documentstyle[epsfig]{mn}

\def\simlt{\mathrel{\rlap{\lower 3pt\hbox{$\sim$}}\raise 2.0pt\hbox{$<$}}}
\def\simgt{\mathrel{\rlap{\lower 3pt\hbox{$\sim$}} \raise 2.0pt\hbox{$>$}}}
\def\lsim{\mathrel{\rlap{\lower 3pt\hbox{$\sim$}}\raise 2.0pt\hbox{$<$}}}
\def\gsim{\mathrel{\rlap{\lower 3pt\hbox{$\sim$}} \raise 2.0pt\hbox{$>$}}}

\def\Zsun{{\rm Z}_{\odot}}

\begin{document}
\title[High redshift GRBs with {\it Swift}]{On the detection of very high redshift Gamma Ray Bursts with {\it Swift}}

\author[Salvaterra et al.]{R.~Salvaterra$^1$, S.~Campana$^2$, G.~Chincarini$^{1,2}$, G.~Tagliaferri$^2$, S.~Covino$^2$ \\
$1$ Dipertimento di Fisica G.~Occhialini, Universit\`a degli Studi di Milano
Bicocca, Piazza della Scienza 3, I-20126 Milano, Italy\\
$2$ INAF, Osservatorio Astronomico di Brera, via E. Bianchi 46, I-23807 Merate
(LC), Italy }

\maketitle \vspace {7cm}

\begin{abstract}
We compute the probability to detect long Gamma Ray Bursts (GRBs) at 
$z\ge 5$ with {\it Swift}, assuming that GRBs form preferentially in 
low--metallicity environments. The model fits 
well both the observed BATSE and {\it Swift} GRB differential peak flux 
distribution and is consistent with the number of $z\ge 2.5$ detections
in the 2--year {\it Swift} data. 
We find that the probability to observe a burst at $z\ge 5$ becomes larger 
than 10\% for photon fluxes $P<1$ ph s$^{-1}$ cm$^{-2}$, consistent with the 
number of confirmed detections. The corresponding fraction of 
$z\ge 5$ bursts in the {\it Swift} catalog is $\sim 10-30$\%  depending on 
the adopted metallicity threshold for GRB formation. 
We propose to use the computed probability as a tool to identify 
high redshift GRBs. By jointly considering 
promptly--available information provided by {\it Swift} and model results, 
we can select reliable $z\ge 5$ candidates in a few hours from the BAT detection.
We test the procedure against last year {\it Swift} data: only three 
bursts match all our requirements, two being confirmed at $z\ge 5$. 
Other three possible candidates are picked up by slightly relaxing the adopted
criteria. No low--$z$ interloper is found among the six candidates.
\end{abstract}


\section{Introduction}

The detection of high redshift objects has been, and is, one of the main 
challenges for Cosmology. Quasars have been since their discoveries the
 beacons of the Universe. Indeed during the last few years the SDSS survey 
started to probe, using high-$z$ quasars, the Universe till near the 
re-ionization epoch (Fan 2006). Long gamma ray bursts (GRB) may 
constitute a 
complementary way to study the cosmos and the early evolution of stars avoiding 
the proximity effects and possibly probing even larger redshifts up to 
$z\sim 10$. The five GRBs detected at $z\gsim 5$, over a sample of about
200 objects observed with the {\it Swift} satellite (Gehrels et al. 2004),
show that a large percentage of GRBs is detected at high-$z$. The current 
record is $z = 6.29$ (Tagliaferri et al. 2005, Kawai et al, 2006). 
One of the goals of the Swift mission is 
the detection of high-$z$ GRBs to estimate at what cosmological epoch they 
exist and, in case they are bright, to use them as beacons of the Universe. 
To achieve this we need to fine tune the BAT instrument decreasing the trigger 
threshold. Moreover, due to the high competition for time on large ground based 
telescopes, we need to preselect the best candidates soon after their detection.

The following analysis is based on the assumption that GRBs form 
preferentially in 
low--metallicity galaxies. The model fits well both the BATSE and 
{\it Swift} GRB differential peak flux distribution and is consistent with 
the number of {\it Swift} identifications at $z\ge 2.5$ and $z\ge 3.5$ in 
the 2--years {\it Swift} data (Salvaterra \& Chincarini 2007; 
thereafter SC07). In this Letter, we compute the probability to detect a 
burst at very high redshift with {\it Swift}  and 
we compare model results with the small number of confirmed detection of 
GRBs at $z\gsim 5$. Finally, we propose to use model results as a tool to 
identify $z\gsim 5$ candidates. By jointly considering theoretical 
predictions and promptly--available information provided by {\it Swift}, 
we can select reliable targets in a few hours from BAT detection. 
We test our selection procedure against last year of {\it Swift} data, 
showing that the method is quite efficient. We propose to use it for
real time selection of high redshift bursts and for dedicated 
follow--up searches of GRB host galaxies at $z\ge 5$.

\section{Model description}

SC07 have computed  the luminosity function (LF) and the formation rate of long 
GRBs by fitting the observed BATSE differential peak flux number 
counts in three different scenarios:  i) GRBs follow the cosmic star 
formation and have a redshift--independent LF; ii) the GRB LF varies with 
redshift; iii) GRBs are associated with star 
formation in low--metallicity environments. In all cases, it is possible to 
obtain a good fit to the data by adjusting the model free parameters.
Using the same LF and formation rate, the models 
reproduce both BATSE and {\it Swift} differential counts, showing 
that the two satellites are observing the same GRB population.
Finally, SC07 have tested the burst redshift 
distribution obtained in the different scenarios against the number of {\it
Swift} detections at $z\ge 2.5$ and $z\ge 3.5$. This procedure allows to 
constrain model results without any assumption on the redshift distribution
of bursts that lack redshift determination and on the effect of 
selection biases (see Fiore et al. 2007 for a detailed discussion about 
this important issue).  Models where GRBs trace the 
star formation rate (SFR) and are described by a constant LF are robustly rule 
out by available data. {\it Swift} high--$z$ detection can be explained 
assuming that the LF is evolving in redshift. In
particular, SC07 found that the typical GRB luminosity should increases with
$(1+z)^\delta$ with $\delta> 1.4$. Alternatively, the large number of 
high--$z$ identifications may indicate that GRBs are biased tracer of the star 
formation, forming preferentially in low--metallicity environment. Assuming 
that the LF does not evolve in redshift, the number of high--$z$ burst 
identifications imply a metallicity threshold for GRBs formation lower than 
$0.3\;\Zsun$ (SC07), consistently with the predictions of collapsar models 
(MacFadyen \& Woosley 1999; Izzard, Ramirez--Ruiz \& Tout 2004).
Available studies of GRB host galaxies (Sollerman et al. 2005, Stanek et
al. 2005, Fynbo et al. 2006b) seem to confirm this metallicity bias. The large 
majority of host galaxies has $Z<0.3\;\Zsun$ although two bursts are 
detected in galaxies with higher metallicity. We want to note here
that many absorption lines detected in optical spectroscopy are possibly 
not probing 
the nearby GRB environment (Jakobsson et al. 2006b, D'Elia et al. 2007, 
Watson et al. 2007, 
Prochaska et al. 2007, Vreeswijk et al. 2007). So, any conclusion about the metallicity of the GRB
progenitors inferred by these studies should be taken with some caution. 

We limit our analysis here to models in which GRBs form preferentially in 
low--metallicity environments and are described by a constant LF, but similar 
results can be obtained 
assuming evolution in the LF. The GRB formation rate is then given by

\begin{equation}
\Psi_{GRB}(z)=k_{GRB} \Sigma(Z_{th},z) \Psi_\star (z),
\end{equation}

\noindent
where $k_{GRB}$ is the GRB formation efficiency, $\Sigma(Z_{th},z)$ gives the
fraction of galaxies at redshift $z$ with metallicity below $Z_{th}$
(Langer \& Norman 2006) and $\Psi_\star(z)$ is the observed star formation 
rate. We adopt the recent determination of the global star formation rate
obtained by Hopkins \& Beacom (2006), slightly modified in order to match
the decline with $(1+z)^{-3.3}$ at $z\ge 5$, as suggested by recent 
deep--field data (see Stark et al. 2007 and references therein). 
In the following we explore a wide range of model with 
$0.02\;\Zsun\le Z_{th}\le 0.3\;\Zsun$, computing the expected probability 
to detect a GRB at $z\ge 5$ with {\it Swift}. For all the 
details of the model computation we refer the interested reader to SC07.

\section{Detection probability at high redshift}

\begin{figure}
\begin{center}
\centerline{\psfig{figure=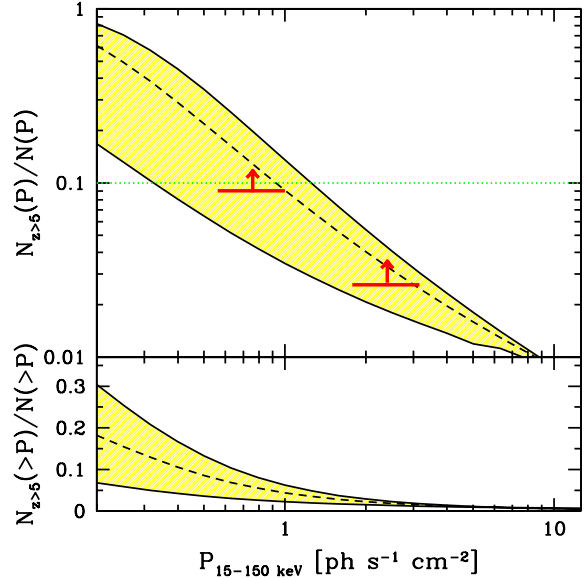,height=8cm}}
\caption{Top panel: probability for a GRB of peak photon flux $P$ to be at $z\ge 5$
for $0.02\;\Zsun\le Z_{th}\le 0.3\;\Zsun$ (shaded area, lower bound refers to higher
metallicity threshold). Dashed line refers to the 
model with $Z_{th}=0.1\;\Zsun$. Horizontal bars refers to lower 
limits derived from the five {\it Swift} $z\gsim 5$ identifications:
GRB 050814 ($z\sim 5.3$, Jakobsson et al. 2006a), GRB 050904 ($z=6.3$, 
Kawai et al. 2006), GRB 060510B ($z=4.9$, Price 2006),
GRB 060522 ($z=5.1$, Cenko et al. 2006), and, GRB 060927 ($z\sim 5.6$,
Fynbo et al. 2006a). Dotted horizontal 
line marks the probability threshold of 10\%. Bottom panel: fraction of
$z\ge 5$ GRBs with photon flux larger than $P$.}
\label{fig:zge5}
\end{center}
\end{figure}

We compute the probability for a GRB of observed peak photon flux $P$ in the
15--150 keV band of {\it Swift}/BAT to be at
$z\ge 5$ as the ratio between the expected number of GRBs at $z\ge 5$ and the 
total number of bursts detectable at peak photon flux $P$. The results are shown
in the top panel of Fig.~\ref{fig:zge5} for 
$0.02\;\Zsun\le Z_{th}\le 0.3\;\Zsun$ (shaded area, lower bound refers to the
higher threshold value). Dotted line refers to the model with 
$Z_{th}=0.1\;\Zsun$. The probability to find a burst at
$z\ge 5$ increases rapidly with decreasing $P$ and becomes larger than 10\% 
for $P<1$ ph s$^{-1}$ cm$^{-2}$. Indeed, four out of five bursts confirmed to
be at $z\gsim 5$ have photon fluxes in the range 0.6--0.8 ph s$^{-1}$ cm$^{-2}$.
Horizontal bars refer to lower limits derived from the available GRB
{\it Swift} detections at $z\gsim 5$ in the corresponding flux bin. 
We find that the model with $Z_{th}=0.1\;\Zsun$ is consistent with 
{\it Swift} lower limits. Models with higher metallicity thresholds tend 
to underestimate the number of detections at $z\gsim 5$. Assuming a threshold
metallicity of $0.3\;\Zsun$, the computed probability falls well below 
available limits. In order to be consistent with observational constraint, we 
have to assume some evolution in the LF. Assuming that GRB luminosity 
increases linearly with redshift, the result of the $Z_{th}=0.3\;\Zsun$ model 
is similar to what obtained with $Z_{th}=0.1\;\Zsun$. Thus, in the following we
will refer as our reference model with a threshold value of $Z_{th}=0.1\;\Zsun$
and no LF evolution, or, as an alternative, with the metallicity threshold of
$0.3\;\Zsun$ and a linear redshift evolution of the burst luminosity.

In the bottom panel of Fig.~\ref{fig:zge5}, we show the cumulative fraction of
GRBs with peak photon flux larger than $P$ expected at $z\ge 5$.
For our reference model, we find that $\sim 18$\% (4\%) of all long 
burst detected by {\it Swift} should lie at $z\ge 5$ assuming a flux limit 
of $P_{lim}=0.2$ (1) ph s$^{-1}$ cm$^{-2}$. Higher (lower) percentages can 
be obtained assuming lower (higher) values of the metallicity threshold. 
Note that the expected fraction of $z\ge 5$ GRBs is higher than the $\sim 10$\%
obtained by Bromm \& Loeb (2006) and 2\% obtained by Gorosabel et al. (2004)
at $P_{lim}=0.2$ ph s$^{-1}$ cm$^{-2}$ for models in which GRBs are tracer of
the cosmic star formation. The discrepancy is easily explained by the 
different intrinsic redshift GRB distributions implied by these models, that 
are found to underestimate also the number of bursts 
detected by {\it Swift} at $z\ge 2.5$ (SC07).
Natarjan et al. (2005) have explored a simple toy model taking into 
account the effect of metallicity in GRB formation. They assumed a higher 
GRB formation rate at $z\ge 3$ when the average metallicity of the
Universe is low and found that the fraction of GRBs at $z\ge 5$ is 
$\sim 30$\%. This value is consistent with our upper bound  obtained 
with a more detailed description of the metallicity redshift evolution.

\begin{figure}
\begin{center}
\centerline{\psfig{figure=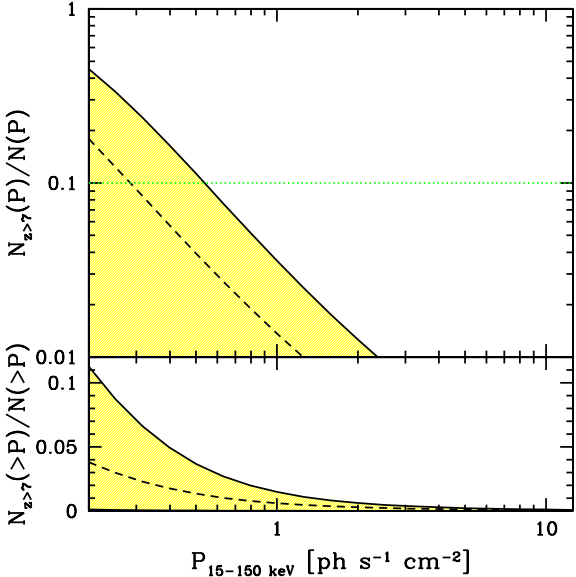,height=8cm}}
\caption{Top panel: probability for a GRB of peak photon flux $P$ to be at $z\ge 7$. 
Bottom panel:  fraction of $z\ge 7$ GRBs with photon flux larger than $P$.
Lines as in the previous figure.}
\label{fig:zge7}
\end{center}
\end{figure}

In Fig.~\ref{fig:zge7}, we show the probability of bursts with observed peak
photon flux $P$ to be detected at $z\ge 7$ (top panel) and corresponding 
cumulative fraction (bottom panel). Up to now, no 
GRB has been identified at so high redshift. Indeed, we find that the 
probability is quite low, but at the lowest photon fluxes. 
For our reference model, the probability is larger than 10\% only for $P<0.3$
ph s$^{-1}$ cm$^{-2}$. Unfortunately, below this limit 
the large majority of the GRBs has no redshift determination. 
We find that $\sim 4$\% (1\%) of all long burst detected by {\it Swift} 
should lie at $z\ge 7$ assuming a flux limit of $P_{lim}=0.2$ (1) ph s$^{-1}$ 
cm$^{-2}$. Daigne, Rossi \& Mochkovitch (2006) computed the expected fraction 
of GRBs at $z\ge 7$ for
a wide range of models considering that GRBs follow the global star formation
but are characterized by strong evolution in the GRB LF. They found that the
expected fraction of $z\ge 7$ GRBs with $P\ge 1$ ph s$^{-1}$ cm$^{-2}$ is 
0.3--4\%, depending on the assumed 
star formation rate, LF evolution and intrinsic peak energy distribution.
Although a full comparison is not possible given the different assumptions in 
the two works, we note that the fraction of $z\ge 7$ GRBs predicted by our 
models (0.1-1.5\%) is consistent with the range of values reported by Daigne 
et al. (2006).

\section{Real time selection of \lowercase{$z>5$} candidates}

Fast selection procedures of high--$z$ GRB candidates are needed in order to
acquire data as the GRB afterglow is still sufficiently bright to allow high 
signal to noise spectroscopic observations and reliable redshift measurements. 
Campana et al. (2007; thereafter C07) proposed to select $z\ge 5$
candidates based on some {\it Swift} promptly--available information: GRB 
time duration ($T_{90}> 60$ s), lack of an UVOT counterpart, and low 
Galactic extinction ($E(B-V)<0.1$ or high galactic latitude). The first constraint 
preferentially selects distant GRBs whose duration is stretched by cosmological
time dilation. The second picks up highly extincted or high redshift objects 
(UVOT observes shortward of 5500 \AA$\;$ and it is virtually blind to objects 
at $z\ge 5$ due to the Lyman drop out). Since March 2006, all bursts are observed 
with the same UVOT strategy: a first short (a few tens of seconds) $V$ image and 
then 100 s white image. So, the requirement of lack of UVOT detection is 
equivalent to have $V>19-21$. Finally, the further constraint on the 
Galactic extinction cleans the sample from heavily absorbed GRBs from our 
Galaxy, i.e. we select bursts with high Galactic latitudes. We note that this 
last requirement is simple operative and not related to the GRB event. 
We further exclude from our selection GRBs that lack an XRT detection
and those for which the satellite did not slew just after the trigger.
Applying the C07 selection criteria on the last year {\it Swift} (from
March 2006 to March 2007), we obtain a small sample of seven candidates.

We propose here to add as further selection criterion the probability of a
burst to be detected at $z\ge 5$, as computed in the previous section. We 
consider as good candidates, GRBs that have probability larger than 10\%, 
corresponding to $P<1$ ph s$^{-1}$ cm$^{-2}$ for our reference model. We 
stress here that the information on the GRB peak photon flux is also promptly 
available in the first GCN 
circulars. Four out of the seven candidates, that satisfy the previous C07
selection criteria, can be indeed discarded on the 
bases of their large photon fluxes: the probability of these GRBs to lie
at $z\ge 5$ is below 2\% in any model here considered. It is interesting to 
note that two out of the four candidates selected in the original work by 
C07 (i.e., GRB 060904A and GRB 060814) do not match our probability criterion. 
Moreover, among the discarded bursts, GRB 070306 is consistent with all
selection criteria considered by C07, being $T_{90}=210$ s, $V>20.5$, and 
Galactic $E(B-V)=0.03$, but it is probably hosted in a galaxy at moderate 
redshift
(Malesani et al. 2007b; Jaunsen et al. 2007 gives $z=1.497$). Indeed, we can 
exclude it from the target sample on the basis of 
the high observed peak photon flux, $P=4.2$ ph s$^{-1}$ cm$^{-2}$, corresponding to 
a probability of $\sim 1-2$\% of being at $z\ge 5$. Therefore, the 
further requirement based on theoretical predictions is very effective in 
reducing the number of possible high redshift candidates, discarding low--$z$ 
interlopers.
By combining model predictions and {\it Swift} promptly available 
information, we select only three bursts out of the 84 GRBs detected 
by {\it Swift} in the last year: GRB 060427, GRB 060510B, and 
GRB 060522. The last two bursts are indeed confirmed to lie at $z\gsim 5$, 
whereas the first one has no redshift. Thus, our method is quite effective in 
selecting $z\ge 5$ GRBs and allow a (near) real time selection of good 
targets for spectroscopic studies of high redshift GRBs with 8-m class 
telescopes. It is important to note that our procedure is very 
effective in single out high-$z$ GRB candidates among the many that are 
detected by {\it Swift}, however it does not guarantee that {\it all} 
high-$z$ bursts will be selected. In other words, it does not select the 
complete sample of GRBs at $z\ge 5$. 

Besides the above GRBs, there are few other {\it Swift} GRBs that match 
all but one of our requirements. We can identify these as secondary candidates. 
Three bursts are picked up by relaxing only slightly our selection criteria: 
GRB 061028, GRB 070129, and GRB 070223. The first two bursts have Galactic
$E(B-V)\sim 0.15$. Since the criterion based on the Galactic extinction 
is only operative cleaning the sample from sources at low Galactic 
latitudes, we can safely include these bursts in our analysis. 
The optical afterglow of 
GRB 070129 has been identified both in $R=21.3$ (Malesani et al. 2007a)
and in $I=20.6$ (Halpern et al. 2007). The detection in these 
bands is not in contrast with a burst lying at $z\ge 5$.
Finally, GRB 070223 is not detected in the UVOT $V$ 
band but the limit in this band is only $>18.9$ mag. 
All these bursts lack of the redshift and might represent good
candidates. So, our
final sample (primary and secondary targets) consists of six bursts, two 
being confirmed at $z\ge 5$ (i.e. success rate $\gsim 33$\%).
Also in this case, the sample is clean, since no low--$z$ 
interlopers has been identified up to now. The properties of the selected 
targets are reported in Table~1.

\begin{table}\tiny
\begin{center}
\begin{tabular}{lcccccl}
\hline
\hline
GRB     & $T_{90}$  &   $P$   &  $V$  & White &  $E(B-V)$  &  $z$ \\
        & [s] & [ph s$^{-1}$ cm$^{-2}$] & & & & \\
\hline
 060427 &   64$\pm$5 &   0.3$\pm$0.1 &   $>20.4$ &   &  0.05 &    \\    
 060510B & 276$\pm$10  &  0.6$\pm$0.1  &  $>21.2$ & $>21.9$ &  0.04 & 4.9      \\
 060522 &   69$\pm$5  &  0.6$\pm$0.2  &  $>20.1$ & $19.7$ &  0.05 & 5.11    \\
\hline
061028 &  106$\pm$5  &  0.7$\pm$0.2 &   $>20.6$ & $>18.9$ &  0.16 &   \\
070129 &  460$\pm$20  &  0.6$\pm$0.1  &  $>20.7$ & $>20.8$ &  0.14 &      \\
070223 & 89$\pm$2 & 0.7$\pm$0.1 & $>18.9$ & $>21.4$ & 0.02 &  \\
\hline
\end{tabular}
\end{center}
\caption{Main properties of the $z\gsim 5$ candidate sample derived from last 
year of {\it Swift} observations. The first three sources are
the one that satisfy all our selection criteria, the other three are just 
slightly below one of them (see text).}
\end{table}

As a further exercise, we apply our selection criteria also to the first
year of {\it Swift} data. Although the UVOT strategy was less efficient
and clean, the procedure gives similar results. Only three bursts satisfy all
our requirements: GRB 050814, GRB 050904, and, GRB 051001. The first two bursts
are indeed at $z\ge 5$, lying at $z\sim 5.3$ and $z=6.3$, respectively, while
the third lacks of redshift determination. This result confirms the efficiency
of the selection procedure here proposed and strengthens our conclusions.

\section{Conclusions}

We compute the probability to detect a long GRB with observed photon flux $P$ 
at very high redshift under the assumption that GRBs form preferentially in 
low--metallicity environments, i.e. are hosted in galaxies with metallicity
below a given threshold. We consider a wide range of metallicity thresholds, 
$0.02\le Z/\Zsun\le 0.3$. The GRB LF and the formation efficiency have been 
obtained by fitting the differential peak flux distribution of bursts detected 
by BATSE. The corresponding redshift distribution is consistent with the 
number of bursts detected at $z\ge 2.5$ and $z\ge 3.5$ in 2--years of 
{\it Swift} mission. We find that a model with $Z_{th}=0.1\;\Zsun$
is consistent with available constraints, whereas higher threshold values
require some luminosity evolution in the GRB LF in order to account for the
five bursts detected up-to-now at $z\gsim 5$. In particular, for 
$Z_{th}=0.3\;\Zsun$, a value consistent with the prediction of some collapsar 
models, the five detections at $z\gsim 5$ can be explained assuming that 
the typical GRB luminosity increases linearly with redshift. 
In this case, we find that the probability to detect 
a GRB at $z\ge 5$ ($z\ge 7$) becomes larger than 10\% for $P< 1$ (0.3) ph 
s$^{-1}$ cm$^{-2}$.
Assuming a flux limit of $0.2$ ph s$^{-1}$ cm$^{-2}$ for {\it Swift}/BAT, 
we expect that 7--30\% (0.1--11\%) of all detected bursts should lie at 
$z\ge 5$ ($z\ge 7$), where the lower (upper) bound refers to the upper (lower) 
metallicity threshold considered. Similar results can be 
obtained by assuming that GRBs are tracer of the global star formation, but 
characterized by strong evolution in the LF (Daigne et al. 2006; SC07). 
However,
these models require that the typical GRB luminosity should increase quite 
strongly with redshift, being $\gsim 7$ times higher at $z=3$ with respect to 
the local one. Moreover, we note that recent observations of GRB host galaxies 
 at high-$z$ (Fynbo et al. 2003) support the idea that GRBs might form
preferentially in low-metallicity environments.

We have seen that the detection of GRBs at high redshift is a rare event. 
Moreover, spectroscopic study of these bursts requires a very rapid 
re--pointing of large ground--based telescope. Therefore, an efficient 
procedure is needed in order to pick--up in real time high--$z$ GRBs.
We propose to use the probability here computed to select reliable $z\ge 5$ 
candidates together with some promptly--available information provided by 
{\it Swift}, such as burst duration, the lack of detection in the UVOT $V$ 
band, and the low Galactic extinction. We test the proposed procedure 
against the data obtained by {\it Swift} in the last year. Three bursts 
match all our requirements,
two being confirmed at $z\gsim 5$ (i.e. success rate $\gsim 66$\%). 
Relaxing slightly our selection criteria, we identify other three
candidates. Therefore, the final sample consists in six bursts (success rate 
$\gsim 33$\%) with no low--redshift interloper identified up to now, showing 
that the proposed procedure is quite efficient in selecting good $z\ge 5$ 
candidates. We want to stress here that all quantities needed for the 
selection are available within a few hours from burst trigger, allowing
a rapid pointing of 8-m class telescope for spectroscopic follow--up
studies. The propose procedure may be also used to identified
reliable targets for dedicated searches of GRB host galaxy at very high 
redshift. 
The detection of large number of very high--$z$ bursts can be a 
fundamental tool to investigate the Universe up to the reionization era.

\end{document}